# Attack based DoS attack detection using multiple classifier

Mohamed Abushwereb
*Department of Computer Science*
*Princess Sumaya University for Technology*
Amman, Jordan
mohamed.abushwereb@gmail.com

Muhannad Mustafa
*Department of Computer Science*
*Princess Sumaya University for Technology*
Amman, Jordan
muhannad.a.mustafa@gmail.com

Mouhammd Al-kasassbeh
*Department of Computer Science*
*Princess Sumaya University for Technology*
Amman, Jordan
m.alkasassbeh@psut.edu.jo

Malik Qasaimeh
*Department of Computer Science*
*Princess Sumaya University for Technology*
Amman, Jordan
m.qasaimeh@psut.edu.jo

**Abstract — **One of the most common internet attacks causing significant economic losses in recent years is the Denial of Service (DoS) flooding attack. As a countermeasure, intrusion detection systems equipped with machine learning classification algorithms were developed to detect anomalies in network traffic. These classification algorithms had varying degrees of success, depending on the type of DoS attack used. In this paper, we use an SNMP-MIB dataset from real testbed to explore the most prominent DoS attacks and the chances of their detection based on the classification algorithm used. The results show that most DOS attacks used nowadays can be detected with high accuracy using machine learning classification techniques based on features provided by SNMP-MIB. We also conclude that of all the attacks we studied, the Slowloris attack had the highest detection rate, on the other hand TCP-SYN had the lowest detection rate throughout all classification techniques, despite being one of the most used DoS attacks.

## I. INTRODUCTION

As the internet expands and assimilates more aspects of our daily life than ever before, from social connections to payments, so has the number of malicious attacks grown in both type and quantity. Therefore, the need to identify these different types of attacks and prevent them from using different network security techniques is paramount [1].

Some attacks aim to gain unauthorized access to systems and information like the buffer overflow and brute force attacks, and other attacks aim to render these systems unavailable altogether such the internet worm and Denial of Service (DoS) flooding attacks [2]. Of all the mentioned types, DoS attacks are infamously known to be the most dangerous and frequently used, causing huge financial losses [3].

To combat these attacks, Intrusion Detection (ID) has become a major component of any networking system. ID aims to detect malicious requests made to the network [4], there are two main approaches to detecting these requests; the first is to compare new requests to previous normal requests and determine if there is an anomaly in the traffic, called Anomaly Intrusion Detection (AID), the second is to compare new requests to known attack signatures and determine if their patterns match, called Misuse Intrusion Detection (MID) [5].

Both of the approaches mentioned above to ID heavily rely on the dataset they use to compare with new requests. For example, when using raw packet data, the processing burden was significant and resulted in slow detection time, this led to the development of different protocols to monitor and manage packet data including Common Management Information Protocol (CMIP), Remote Network Monitoring (RNM), and Simple Network Management Protocol (SNMP) [6].

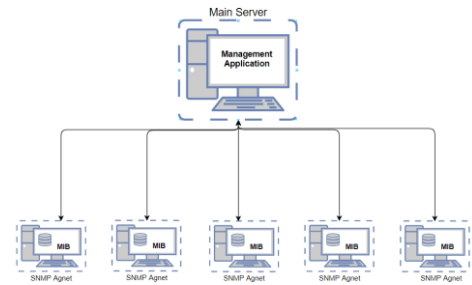

Fig. 1. Network Management Architecture.

SNMP gathers information from all the different devices connected to a network, not only is this information from different devices but is also collected across different networking and protocol layers (TCP, UDP, IP, ICMP, etc) [7]. Fig. 1 shows how SNMP coupled with MIB is one of the fundamentals when it comes to any Internet management model.

To make use of this large amount of data, a database system known as a Management Information Base (MIB) uses several management variables to store this data in a statistical format when coupling MIB to SNMP, the benefits of passive monitoring, collection, and organized storage of data provides a solid ground for characterizing and detecting all types of different attacks [5].

In this work, the aim is to analyze the most used DOS attacks individually and compare them in respect to detection rates using the dataset proposed in [8]. The paper is structured as follows: First, section 2 reviews several relevant papers that have used machine learning in intrusion detection systems. Then, section 3 goes through the methodology of the paper, where an overview of DOS attacks, the SNMP-MIB dataset, and Machine Learning classifiers is given, ended with the details of implementation. In section 4, the results of the experiment and observations drawn from these results are displayed. Finally, the conclusions of this research paper are presented in section 5.

## II. RELATED WORK

Since detecting network anomalies and intrusions is extremely beneficial, many studies and methods were implemented and tested in the past [9]. Most of these approaches focused on analyzing raw traffic data based on network flow, IP addresses, number of packets, ports, and other key features. Another type of approach to anomaly detection uses the Simple Network Management Protocol Management Information Base (SNMP MIB) as a source for data.

This section explores previous research centered around the second approach, intrusion detection using SNMP-MIB. One of the early attempts in this approach was made in [10], where the authors used statistical tests for casualty and derived different MIB variables in aims of early detection of Distributed Denial of Service (DDoS) attacks. To derive these variables, three different types of DDoS attacks were conducted, and 91 MIB variables of five different groups (TCP, UDP, IP, ICMP, and SNMP) were monitored in both the attacker and target. The evaluation of this approach shows that it is possible to predict and detect a DDoS attack before the target shuts down with an approximate false alarm rate of only 1%.

[11] managed to detect flooding attacks with a rate of 93.0% using SNMP-MIB parameters and a C4.5 algorithm. These parameters were based on the simulation environment operating the flooding attack and were fed to a custom C4.5 algorithm that used the parameters to classify different types of traffic and detect flooding attacks.

[12] also used SNMP-MIB parameters to detect a type of attack, namely the ARP spoofing attack. They experiment with three different approaches; a Naive Bayesian algorithm, a support vector machine algorithm, and finally a C4.5 algorithm. These approaches were each tested and then compared. False-positive rate, accuracy rate, and missing rate were metrics used to identify each approach's strengths as well as weaknesses. The conclusion was that the C4.5 algorithm had the highest accuracy rate, while the lowest accuracy rate belonged to the Naive Bayesian algorithm, and the support vector machine algorithm had the lowest number of false alarms.

On the other hand, the authors in [13] used a two-phased decentralized detection mechanism based on the clustering algorithm. The first phase was a monitoring phase in which several sources were monitored to collect SNMP-MIB parameters. The second phase was the traffic detection phase which used the SNMP-MIB parameters to classify the traffic into abnormal and normal traffic. To test this decentralized detection mechanism, Cerroni et al used a subset of the SNMP-MIB dataset, that which concerns decentralized detection mechanisms, and when tested, showed that the proposed two-phase algorithm was able to successfully identify plausible intrusions in that subset of the SNMP-MIB.

[14] also devised a new distributed data mining method. This method used fourteen SNMP-MIB parameters (related to the IP and TCP groups) based on a simulated network environment to detect a specific type of DDOS attack. The method was evaluated for decentralized testbed environments and had a satisfactory detection rate.

In the work of [15], another SNMP MIB based early detection approach was proposed, this time using statistical tests for causality. The MIB variables were collected by sampling network traffic from four groups; TCP, UDP, IP, and ICMP. The sampling took place every five seconds. This approach was tested with five types of DDoS attacks and demonstrated a high accuracy rate in detection.

The research [16] studies five MIB variables in accordance with IF and IP groups; these variables were used to derive a time series that is in turn input for an Auto-Regressive (AR) algorithm. This method was then tested using real instances of UDP, SYN, Smurf, and ICMP flood attacks, the results suggesting that the attacks were detected successfully.

On a similar note, paper [17] studied four MIB variables based on the IF group, these variables were also used by an anomaly detection algorithm. The approach was tested against two real instances of SYN, Smurf, and DOS flood attacks, and was able to efficiently discover them.

In paper [18], SNMP MIB variable correlation was used for attack detection, the variables totaled to 16 and gathered from 6 groups. These variables were used by a lightweight algorithm that was tested against UDP, ICMP, and TCP-SYN flooding attacks, the algorithms shown to accurately detect all attacks with minimal false negative and false-positive rates.

Work of [2] and [19] proposed an intrusion detection system that classifies a flooding attack using an SVM machine learning approach. The system uses 13 SNMP MIB variables sampled from 4 MIB groups at a rate of 15 seconds. The system used an SVM hierarchal structure to classify traffic into three different types of attacks; ICMP, UDP, and TCP-SYN flood attacks. This system when tested showed to have an accuracy rate of 99.27% when detecting known and also new attacks.

The authors in paper [20] proposed a system that could predict a Trojan attack using a decision tree algorithm. This algorithm used SNMP MIB data, primarily MIB variables connected to Host Resources such as running services and software, because of how these variables can indicate changes in a system's performance.

In paper [21] SNMP MIB and Machine Learning were used to build a three modeled intrusion detection system. The first module uses the C4.5, attribute selection, and RIPPER algorithms to select the top MIB variables, these variables were in turn used by the second model to build an intrusion detection system, which the third model uses for real-time DOS detection. The system was tested using four groups in the MIB related to ICMP, UDP, and TCP-SYN attacks and gained a high detection rate of 99.03%

Paper [22] introduces a system that detects special kinds of DOS attacks such as DNS and TCP SYN reflection flooding attacks. The system was named Protocol Independent Detection and Classification (PIDC) and used an algorithm that derived the correlation between 13 MIB variables and ranked them. These variables were collected from the DNS and TCP groups. Then a C4.5 algorithm classified the attacks used in the research using the ranked variables. This system detected the attacks with 1% false-positive rates and 99% true positive rates.

Statistical signal processing was used in paper [23], along with three MIB variables form the Internet Protocol group. The authors used case studies such as Network Access Problems and Server Failure to demonstrate a hypothesis that is based on the Generalized Likelihood Ratio (GLR), where correlations are made in accordance to changes in the MIB variable's behavior. Results concluded that statistical signal processing has big potential in intrusion detection.

Authors in [24] and [25] aimed to bypass the limitations found in centralized intrusion detection systems, they did this using a Mobile Agent that collects local machine's MIB variables. They then combined the Wiener filter statistical method and MA technology to detect intrusions by inputting MIB variables and desired MIB variables. The MIB variables were from the IP and IF groups. When tested against brute force attacks, null session attacks, buffer overflow, and decoy port-scan scenarios, the model detected all attacks in light traffic networks, while having difficulties in detecting the same attacks in high traffic networks.

Paper [26] proposed an anomaly detection method that uses information entropy and the SNMP MIB dataset. This method was tested on four case scenarios; application server shutdown failure, database failure, 100% cup occupancy, and database shutdown failure.

It is noticeable that while a large number of works have studied DOS attack detection using machine learning classification and layer-based approaches, even enhancing these approaches with SNMP-MIB variables, none of the research made -to our knowledge- has investigated these affairs from an attack based perspective. This paper analyses how each attack is reacted to by machine learning classifiers, and attempts to find the least detected attacks and compares the results found to real-world data of instances of malicious DOS attacks.

III. DENIAL OF SERVICE ATTACK AND SNMP-MIB DATASET

*A. DOS Attack*

There are many methods used in compromising networks, from eavesdropping to phishing to SQL injections. However one of the most used methods by attackers is the Denial of Service attack (DOS). Where other attacks might aim to obtain unauthorized access or protected data, DOS attacks aim to shut down the network or service it provides. As defined by NIST, a DOS attack is "A set of actions that comprise networks and its resources and preventing the authorized users from doing their intended functions".

To achieve this function, a DOS attack has a number of resources it can target and compromise; first is network bandwidth, where the attack targets the connection between a web server and the appliances that connect to it, or the attack targets the connection between the web server and the global internet. The second resource a DOS attack can target is the system's resources, where it overloads the system with continuous requests impeding it from responding to actual authorized users. The third and final target for a DOS attack is application resources, where the attack does not overload the whole system, only a specific application of that system, rendering it unavailable.

A number of papers have proposed approaches to detect different types of DOS attacks based on machine learning classifiers, in this paper we take a look at these different intrusion detection methods from an attack focused perspective, aiming to glean insights on each attack's effectiveness and the threat it poses when guarded against by the most known and used intrusion detection methods today.

These are the types of DOS attacks that are most used by attackers:

SYN Spoofing: this is a basic flooding attack targeting the servers' tables that manage connections to clients. It exploits the three-way handshake protocol used in TCP connection requests. A normal three-way handshake procedure starts with a client sending a TCP SYN request, to which the server replies with an SYN-ACK, finally the client sends an ACK packet confirming the connection.

What the SYN spoofing attack does is send many TCP SYN requests to the server, only with counterfeit source IP addresses, to which the server sends SYN-ACK requests. This causes two effects; the first is that the large number TCP SYN requests are stored in the table without ever receiving the proper ACK packets because the counterfeit IP addresses don't reply to the SYN-ACK requests the server sends out. Second is that For every SYN-ACK the server sends, it reserves a spot on the table for the supposed ACK confirmation. This results in the server's table being filled with both incomplete TCP SYN and SYN-ACK requests and unable to fulfil other client's connection requests.

UDP Flooding: this attack targets a specific port in the victim system or server, it then floods that port with UDP packets to overload the port and shut down the service it provides.

HTTP Flooding: an attack based on HTTP protocols, achieved by having several bots flooding the web server with tasking HTTP requests that deplete its resources like a memory for example.

Slowloris: another HTTP based attack, here the attacker establishes several connections to the server based on incomplete HTTP requests, to keep these connections from being terminated, continuous header lines are sent to keep the connection alive, this exhausts all the connections a server can establish causing it to deny service to any other connection.

Slowpost: close to the Slowloris attack, Slowpost is another HTTP based attack that sends tasking HTTP requests to the victim server, the difference here is that the HTTP requests are complete, with the content-length field defined signalling a posted message to the request; however the content is delivered very slowly with a rate of one byte every two minutes, ultimately overloading the server with unfinished HTTP requests.

ICMP Echo: echo requests are used regularly for diagnostics and are important for TCP/IP networks, attackers flood the server with echo requests and ICMP packets making it hard for any client to reach the server effectively. System administrators usually restrict this protocol with firewalls or intrusion detection systems.

Brute force: this attack is known to be used as a naive approach to password bypassing, but it is also used as a DOS attack, where the server is flooded with password authentication requests and its resources are depleted.

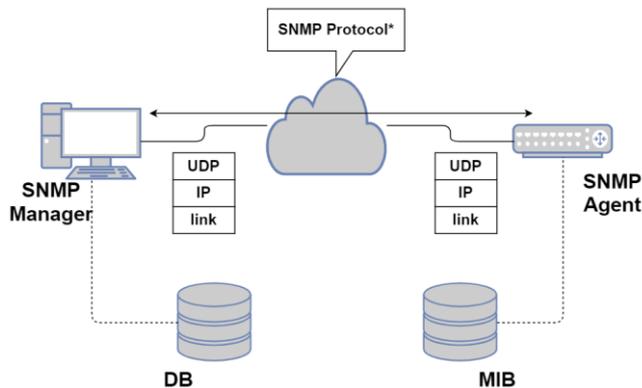

Fig. 2. Network Managemnt Architec

Table I shows how the dataset's records are classified in relation to the attacks mentioned above

TABLE I. DATASET RECORDS ACCORDING TO RELATED ATTACKS

| No. | Traffic Label | Traffic Count |
|---|---|---|
| 1 | Normal | 600 |
| 2 | ICMP-Echo Attack | 632 |
| 3 | TCP-SYN Attack | 960 |
| 4 | UDP Flood Attack | 773 |
| 5 | HTTP Flood Attack | 573 |
| 6 | Slowloris Attack | 780 |
| 7 | Slowpost Attack | 480 |
| 8 | Brute Force Attack | 200 |

### B. Simple Network Management Protocol (SNMP)

To better control monitor and analyze traffic within a certain network, an application layer protocol called Simple Network Management Protocol (SNMP) was developed. SNMP consists of two subsystems as shown in Fig.2, an SNMP Agent and SNMP Manager. The SNMP Agent is embedded in the appliances we want to monitor, and in turn, it collects data from that appliance's network and stores it in a Management Information Base (MIB), a type of database. The SNMP Manager requests that data from the SNMP Agent and uses it to provide a number of network management tasks such as configuration, security measures, and monitoring performance and faults.

There are three versions of SNMP, each with different features; SNMP community is divided into SNMPv1 and SNMPv2, while SNMPv3 is the version designed with extra security features and thus dubbed SNMP security.

### C. Machine learning

Machine learning is a method used to predict and label information, when a machine learning technique uses an already labeled dataset to train the algorithm before labeling the target dataset, it is called supervised machine learning. Classification is a type of supervised machine learning technique, its high accuracy rate leads it to being used in a number of Intrusion Detection systems to classify traffic into normal and abnormal traffic with a good degree of success. This is a list of the machine learning classifiers used in this experiment: BayesNET, NaiveBayes, J48, Random Forest, LMT, Random Tree, NaiveBayesUpdatable, Logistic, Simple Logistic, SMO, MultilayerPerception, IBK, and MultiClassClassifier.

Random Forest is a machine learning classifier made up of a number of decision trees that operate as a group, where the most voted prediction is accepted[27].

Table II shows the two classifiers with the best results, random forest and logistic

TABLE II. TOP TWO CLASSIFIERS RESULTS

| Class | Random Forest | Logistic |
|---|---|---|
| Normal | 100% | 100% |
| ICMP-Echo Attack | 100% | 100% |
| TCP-SYN Attack | 100% | 99.9% |
| UDP Flood Attack | 100% | 100% |
| HTTP Flood Attack | 100% | 99.9% |
| Slowloris Attack | 100% | 99.9% |
| Slowpost Attack | 100% | 100% |
| Brute Force Attack | 100% | 100% |

### D. Implmntion

The experiment implemented in this work aims to test a number of common DOS flooding attacks against the Machine Learning classifiers used in intrusion detection systems. This was done using the MIB dataset, from which 70% (3498 records) was randomly selected as a training dataset for the classifiers, and the remaining 30% (1500 records) was used to testing the trained classifiers. Both the training and testing datasets had instances of normal and the seven attacks previewed in the DOS section. The experiment used Interface MIB variables before applying classification algorithms to the set.

This was done on a device running Windows Server 2016 Datacenter, a 64-bit Operating System, with an Intel(R) Xeon(R) E5-2673 v4 processor and 16 GB of memory (RAM). The Machine Learning algorithms were sourced from Weka 3.8, an open source collection of Machine learning tools and algorithms.

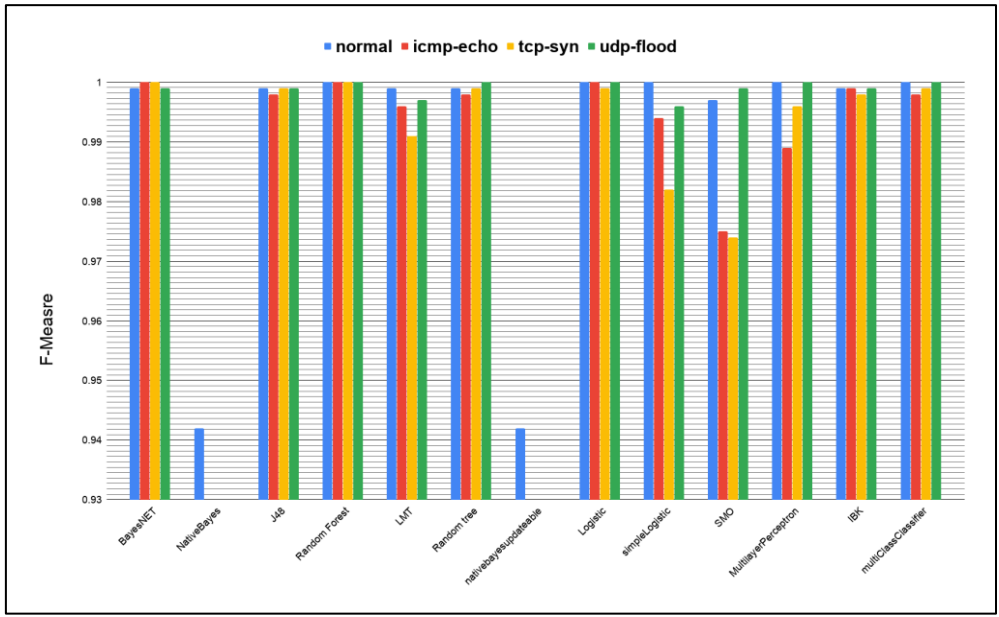

Fig. 4. F-Measure for Network Based DoS attacks

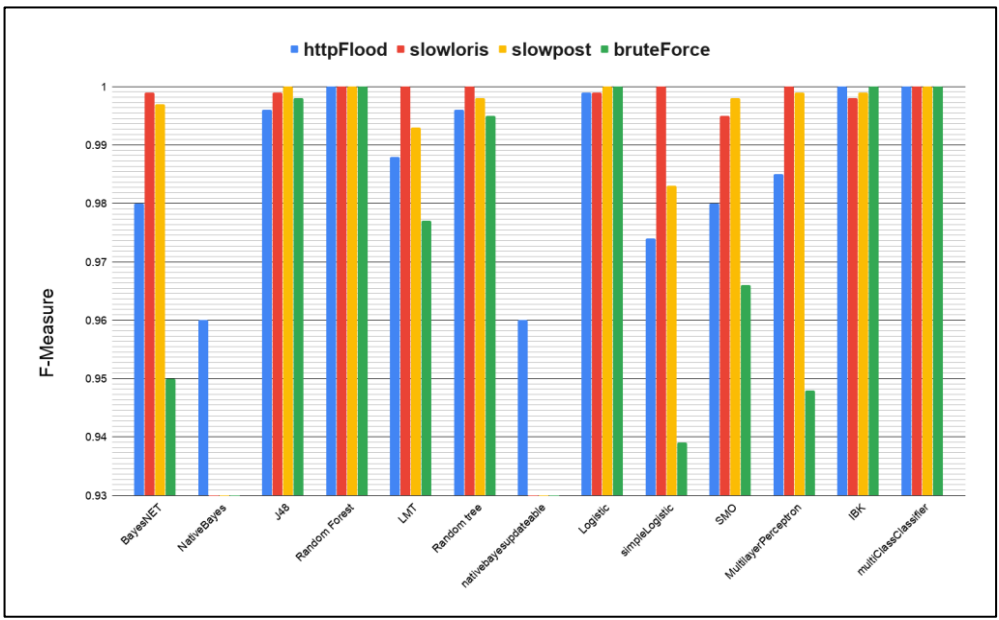

Fig. 3. F-Measure for Web Based DoS attacks

## IV. RESULTS AND DISCUSSION

To better make sense of the results of this experiment, we sorted the attacks studied into either network-based attacks or web-based attacks. Also, the results were filtered to include only a detection rate of 93% or above, considering any lower detection rate not an adequate security measure and therefore out of this paper's scope.

### A. Algorithms act

In this section, we will first take a comparative look at the efficiency of each algorithm and how they respond to two broad groups of attacks; network-based attacks that consist of normal, ICMP-Echo, TCP-SYN, and UDP Flood attacks, and Web based attacks that include HTTP Flood, Slowloris, Slowpost, and Brute force attacks.

Secondly, we will take a more in-depth look into each attack individually and how sensitive they were to detection.

*a) Network Based Attack*

Fig.3. shows how various algorithms respond to three network-based attacks with a detection rate of 93% or more. It is noticeable that NaiveBayes and NaiveBayesUpdatable algorithms did not detect network-based attacks (other than normal) with a rate more than 93%. From an attack centered perspective, out of 13 algorithms, 11 were able to detect normal attacks with a high rate of 99.7% or more, while only 7 algorithms were able to achieve the same detection rate for TCP-SYN attacks.

From an algorithm centered perspective, only Random Forest and Logistic algorithms successfully detected both UDP flood and ICMP echo with 100% accuracy, it is also noted that out of all the algorithms, only Random Forest detected all network-based attacks.

*b) Web based Attack*

Fig.4. shows how the same algorithms in the previous chart respond to four web-based attacks with a detection rate of 93% or more. Firstly, both MultiClassClassifier and Random Forest algorithms detected 100% of all web-based attacks, on the other hand, NaiveBayes and NaiveBayesUpdatable algorithms only detected HTTP Flood out of the four attacks, and even then with a rate of only 96%.

Finally, from an attack based perspective, Slowloris attacks had the highest number of detections with 6 algorithms achieving a full 100% detection rate, while Brute Force attacks had the lowest number of detections with 6 algorithms getting a detection rate lower than 97%.

*B. Attack Based Sensitivity*

In Fig.5. , we explore the various attacks presented in this paper based on the number of algorithms that detected it with 100% accuracy. We notice that the highest sensitivity belonged to Slowloris, with 6 algorithms out of 13 completely detecting the attack, while the lowest sensitivity belonged to TCP-SYN, with only 2 algorithms out of 13 achieving 100% detection rates. This is due to the attack packets used in SYN flooding, as they are not different from normal packets except for the spoofed address source, therefore it is difficult to distinguish them from any normal packet intended for the victim server, making SYN flooding attacks difficult to detect.

Attesting to the low sensitivity displayed by algorithms towards TCP-SYN attacks in our experiment, Fig.6. sourced from Kaspersky shows the attacks most used when targeting networks and systems, with SYN type attacks being notably responsible for 79.70% of all DoS attacks on networks.

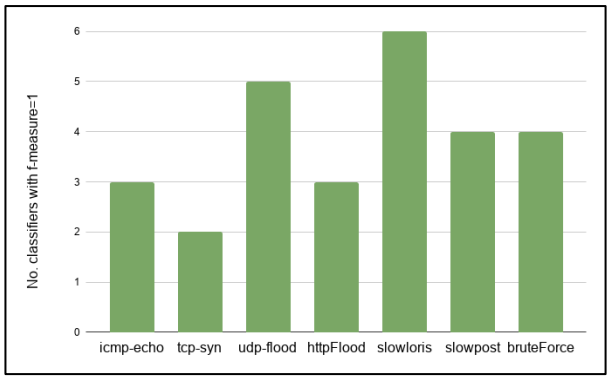

Fig. 6.  Attack-Based Sensitivity

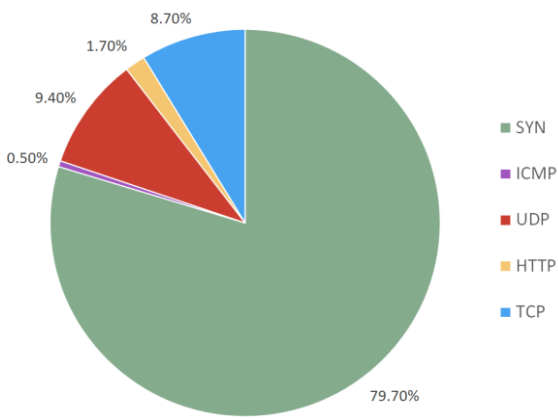

Fig. 5.  Distribution of DoS attacks by type in Q3 2019 [28]

## V. CONCLUSION

In this work, the aim was to study important DOS attacks and how they interact with different machine learning classification methods. We concluded that most DOS attacks are detected with a high accuracy rate of 93% or more when using machine learning classification techniques with an SNMP-MIB features.

The Slowloris attack showed the highest sensitivity to detection, with 6 classification algorithms detecting 100% of the attacks. On the other hand, only 2 algorithms completely detected the TCP-SYN attacks, due to the small difference malicious flood SYN packets have compared to normal SYN packets. These results related to TCP-SYN correlate to Kaspersky reports in which TCP-SYN has the highest number throughout all DOS attack instances, with a total of 79.70%.

We conclude that the TCP-SYN attack is a good target for future study, either through building a large dataset dedicated to TCP-SYN through use of multiple tools and sources, or through exploring other approaches to identifying and detecting TCP-SYN attacks.